\documentclass[12pt]{article}
																																																																																											\pdfoutput=1																																																																																			 
\usepackage{amsmath,amssymb}
\usepackage{graphicx}
\usepackage[pdftex]{hyperref}
\DeclareGraphicsExtensions{.eps,.bmp,.wmf,.jpeg,.pdf}
\numberwithin{equation}{section}
\def\be{\begin{equation}}
\def\ee{\end{equation}}

\def\bea{\begin{eqnarray}}
\def\eea{\end{eqnarray}}
\title{Quintessential and phantom power-law solutions in scalar tensor model of dark energy}
\author{L. N. Granda\thanks{ngranda@univalle.edu.co}\, \ D.F. Jimenez \thanks{jimenez.diego@correounivalle.edu.co}, and\ C. Sanchez \thanks{cristhian.sanchez@correounivalle.edu.co}\\ {\small\it Departamento de Fisica, Universidad del Valle}\\{\small\it A.A. 25360, Cali, Colombia}}
\date{}
\begin{document}
\maketitle

\begin{abstract}
\noindent  We consider a scalar-tensor model of dark energy with kinetic and Gauss Bonnet couplings. We study the conditions for the existence of quintessential and phantom power-law expansion, and also analyze these conditions in absence of potential (closely related to string theory). A mechanism to avoid the Big Rip singularity in various asymptotic limits of the model has been studied. It was found that the kinetic and Gauss-Bonnet couplings might prevent the Big Rip singularity in a phantom scenario. The autonomous system for the model has been used to study the stability properties of the power-law solution, and the centre manifold analysis was used to treat zero eigenvalues.\\ 

\end{abstract}

\section{Introduction}
\noindent 
According to recent observations the current universe undergoes a phase of accelerated expansion, due to domination of dark energy (DE) over the matter content of the universe(usual barionic and dark  matter) \cite{riess,perlmutter,kowalski,hicken,komatsu,percival}. The astrophysical observations indicate that the equation of state parameter for dark energy lies in a narrow region around $w =-1$, which might include values smaller than $-1$. The current observational data are in agreement with the simplest possibility of the cosmological constant as the source of DE, but there is no mechanism to explain its smallness (expressed in Planck units) in contradiction with the expected value as the vacuum energy in particle physics \cite{peebles}, \cite{padmana1}; and observational data also show a better fit for a redshift dependent equation of state. Despite the variety of DE models, it is however difficult to fulfill all observational requirements like the observed value of the equation of state parameter (EoS) of DE, $w\approx -1$, the current content of DE relative to that of dark matter (known as coincidence problem), and the estimated redshift transition between decelerated and accelerated phases, among others. 
Among the models used to explain the DE (for review see \cite{peebles, padmana1, sahnii, copeland, sergei11, odintsovsd}), the scalar-tensor theories are some of the most studied, not only because they contain direct couplings of the scalar field to the curvature, many of them predicted by fundamental theories like Kaluza-Klein and string theories \cite{peri}, \cite{maeda}, but also because the scalar-tensor models fulfill in principle many of the above requirements.\\
In the present work we consider a string and higher-dimensional gravity inspired scalar-tensor model, with non minimal kinetic and Gauss Bonnet (GB) couplings, to study late time cosmological dynamics and board some issues of dark energy. These terms are present in the next to leading $\alpha'$ corrections in the string effective action (where the coupling coefficients are functions of the scalar field) \cite{tseytlin}, \cite{meissner1} and have the notorious advantage that lead to second order differential equations, preserving the theory ghost free. \\
Some late time cosmological aspects of scalar field model with derivative couplings to curvature have been considered in \cite{sushkov}, \cite{granda,granda1,granda2}, \cite{gao}, \cite{granda3}. On the other hand, the GB invariant coupled to scalar field has been extensively studied. In \cite{sergei12} the GB correction was proposed to study the dynamics of dark energy, where it was found that quintessence or phantom phase may occur in the late time universe. Different aspects of accelerating cosmologies with GB correction have been also discussed in \cite{tsujikawa}, \cite{leith}, \cite{koivisto1}, \cite{koivisto2}, \cite{neupane}, \cite{bamba1}. The modified GB theory applied to dark energy has been suggested in \cite{sergei14}, and further studies of the modified GB model applied to late time acceleration, have been considered among others, in \cite{sergeio1}, \cite{sergeio2}, \cite{sergei15}, \cite{carter}, \cite{tretyakov}. The combined effect of GB and kinetic coupling to curvature in the context of dark energy, has been considered in \cite{granda10,granda11,granda12,granda13}. In \cite{granda10} solutions with Big Rip and Little Rip singularities have been considered, in \cite{granda11,granda12} the reconstruction of different cosmological scenarios, including known phenomenological models has been studied, and in \cite{granda13} some exact solutions have been found.\\
In the present paper we focus on the power-law solutions of the quintessence and phantom types, in the case of late-time cosmology with scalar field dominance. We analyze the conditions to avoid the Big Rip singularity presented in phantom power-law solutions. The autonomous system for the model has been considered to find the restrictions on the parameter space of the model satisfying the conditions of stability.
In section II we introduce the model and give the general equations, which are then expanded on the FRW metric to study the different power-law solutions. In section III we analyze the conditions to evade the Big Rip singularity in different scenarios. In section IV we introduce the dynamical variables and analyze the stability of the power-law solutions. Concluding remarks are given in section V.
\section{The model and power-law solutions}
We consider the following action that contains the Gauss Bonnet coupling to the scalar field and kinetic couplings to curvature (such terms are present in the leading $\alpha'$ correction to the string effective action \cite{meissner1}). 
\be\label{eq1}
\begin{aligned}
S=&\int d^{4}x\sqrt{-g}\Big[\frac{1}{16\pi G} R -\frac{1}{2}\gamma\partial_{\mu}\phi\partial^{\mu}\phi+F_1(\phi)G_{\mu\nu}\partial^{\mu}\phi\partial^{\nu}\phi- V(\phi)-F_2(\phi){\cal G}\Big]
\end{aligned}
\ee
\noindent where $\gamma=\pm 1$ ($+1$ for the standard scalar field and $-1$ for the phantom scalar), $G_{\mu\nu}=R_{\mu\nu}-\frac{1}{2}g_{\mu\nu}R$, ${\cal G}$ is the 4-dimensional GB invariant ${\cal G}=R^2-4R_{\mu\nu}R^{\mu\nu}+R_{\mu\nu\rho\sigma}R^{\mu\nu\rho\sigma}$. The coupling $F_1(\phi)$ has dimension of $(length)^2$, and the coupling $F_2(\phi)$ is dimensionless. 
Note that compared with the more general action that leads to the second-order equations of motion (in metric and scalar field) \cite{cartier}, we are neglecting derivative terms that are not directly coupled to curvature, of the form $\Box\phi\partial_{\mu}\phi\partial^{\mu}\phi$ and $(\partial_{\mu}\phi\partial^{\mu}\phi)^2$, which is is acceptable in a cosmological scenario with accelerated expansion. The properties of the GB invariant guarantee the absence of ghost terms in the theory. Hence, the equations derived from this action contain only second derivatives of the metric and the scalar field.\\
By varying Eq. (\ref{eq1}) with respect to metric we derive the gravitational field equations given by the expressions
\be\label{eq2}
R_{\mu\nu}-\frac{1}{2}g_{\mu\nu}R=\kappa^2 \left(T_{\mu\nu}+T^{(m)}_{\mu\nu}\right)
\ee
where $\kappa^2=8\pi G$, $T_{\mu\nu}^m$ is the usual energy-momentum tensor for the matter component, the tensor $T_{\mu\nu}$ represents the variation of the terms which depend on the scalar field $\phi$ and can be written as
\be\label{eq3}
T_{\mu\nu}=T_{\mu\nu}^{\phi}+T_{\mu\nu}^{K}+T_{\mu\nu}^{GB}
\ee
where $T_{\mu\nu}^{\phi}$ corresponds to the variations of the standard minimally coupled terms, $T_{\mu\nu}^{K}$ comes from the kinetic coupling, and $T_{\mu\nu}^{GB}$ comes from the variation of the coupling with GB. 
Due to the kinetic coupling with curvature and the GB coupling, the quantities derived from this energy-momentum tensors will be considered as effective ones. The respective components of the energy-momentum tensor (\ref{eq3}) are given by 
\be\label{eq4}
T_{\mu\nu}^{\phi}=\gamma\nabla_{\mu}\phi\nabla_{\nu}\phi-\frac{1}{2}\gamma g_{\mu\nu}\nabla_{\lambda}\phi\nabla^{\lambda}\phi
-g_{\mu\nu}V(\phi)
\ee
\be\label{eq5}
\begin{aligned}
T_{\mu\nu}^{K}=&\left(R_{\mu\nu}-\frac{1}{2}g_{\mu\nu}R\right)F_1(\phi)\nabla_{\lambda}\phi\nabla^{\lambda}\phi+g_{\mu\nu}\nabla_{\lambda}\nabla^{\lambda}\left(F_1(\phi)\nabla_{\gamma}\phi\nabla^{\gamma}\phi\right)\\
&-\frac{1}{2}(\nabla_{\mu}\nabla_{\nu}+\nabla_{\nu}\nabla_{\mu})\left(F_1(\phi)\nabla_{\lambda}\phi\nabla^{\lambda}\phi\right)+R F_1(\phi)\nabla_{\mu}\phi\nabla_{\nu}\phi\\& -2F_1(\phi)\left(R_{\mu\lambda}\nabla^{\lambda}\phi\nabla_{\nu}\phi+R_{\nu\lambda}\nabla^{\lambda}\phi\nabla_{\mu}\phi\right)+g_{\mu\nu}R_{\lambda\gamma}F_1(\phi)\nabla^{\lambda}\phi\nabla^{\gamma}\phi\\
&+\nabla_{\lambda}\nabla_{\mu}\left(F_1(\phi)\nabla^{\lambda}\phi\nabla_{\nu}\phi\right)+\nabla_{\lambda}\nabla_{\nu}\left(F_1(\phi)\nabla^{\lambda}\phi\nabla_{\mu}\phi\right)\\
&-\nabla_{\lambda}\nabla^{\lambda}\left(F_1(\phi)\nabla_{\mu}\phi\nabla_{\nu}\phi\right)-g_{\mu\nu}\nabla_{\lambda}\nabla_{\gamma}\left(F_1(\phi)\nabla^{\lambda}\phi\nabla^{\gamma}\phi\right)
\end{aligned}
\ee
and 
\be\label{eq6}
\begin{aligned}
T_{\mu\nu}^{GB}=&4\Big([\nabla_{\mu}\nabla_{\nu}F_2(\phi)]R-g_{\mu\nu}[\nabla_{\rho}\nabla^{\rho}F_2(\phi)]R-2[\nabla^{\rho}\nabla_{\mu}F_2(\phi)]R_{\nu\rho}-2[\nabla^{\rho}\nabla_{\nu}F_2(\phi)]R_{\nu\rho}\\
&+2[\nabla_{\rho}\nabla^{\rho}F_2(\phi)]R_{\mu\nu}+2g_{\mu\nu}[\nabla^{\rho}\nabla^{\sigma}F_2(\phi)]R_{\rho\sigma}-2[\nabla^{\rho}\nabla^{\sigma}F_2(\phi)]R_{\mu\rho\nu\sigma}\Big)
\end{aligned}
\ee
In this last expression the properties of the 4-dimensional GB invariant have been used (see \cite{farhoudi}, \cite{sergeiod}).\\
By varying with respect to the scalar field gives us the equation of motion
\be\label{eq7}
\begin{aligned}
&-\frac{1}{\sqrt{-g}}\partial_{\mu}\left[\sqrt{-g}\left(R F_1(\phi)\partial^{\mu}\phi-2R^{\mu\nu}F_1(\phi)\partial_{\nu}\phi+\gamma\partial^{\mu}\phi\right)\right]+\frac{dV}{d\phi}+\\
&\frac{dF_1}{d\phi}\left(R\partial_{\mu}\phi\partial^{\mu}\phi-2 R_{\mu\nu}\partial^{\mu}\phi\partial^{\nu}\phi\right)-\frac{dF_2}{d\phi}{\cal G}=0
\end{aligned}
\ee
Let us consider the spatially-flat Friedmann-Robertson-Walker (FRW) metric,
\be\label{eq8}
ds^2=-dt^2+a(t)^2\left(dr^2+r^2d\Omega^2\right)
\ee
where $a(t)$ is the scale factor. Replacing this metric in Eqs. (\ref{eq2})-(\ref{eq8}) we obtain the set of equations describing the dynamical evolution of the FRW background and the scalar field in the present model:
\be\label{eq9}
H^2=\frac{\kappa^2}{3}\rho_{eff}=\frac{\kappa^2}{3}\left(\frac{1}{2}\gamma\dot{\phi}^2+V(\phi)+9 H^2F_1(\phi)\dot{\phi}^2+24H^3\frac{dF_2}{d\phi}\dot{\phi}\right)
\ee
\be\label{eq10}
\begin{aligned}
&-2\dot{H}-3H^2=\kappa^2 p_{eff}=\kappa^2\Big[\frac{1}{2}\gamma\dot{\phi}^2-V(\phi)-\left(3H^2+2\dot{H}\right)F_1(\phi)\dot{\phi}^2\\&-2 H\left(2F_1(\phi)\dot{\phi}\ddot{\phi}+\frac{dF_1}{d\phi}\dot{\phi}^3\right)
-8H^2\frac{dF_2}{d\phi}\ddot{\phi}-8H^2\frac{d^2F_2}{d\phi^2}\dot{\phi}^2-16H\dot{H}\frac{dF_2}{d\phi}\dot{\phi}-16H^3\frac{dF_2}{d\phi}\dot{\phi}\Big]
\end{aligned}
\ee

\be\label{eq13}
\begin{aligned}
&\gamma\ddot{\phi}+3\gamma H\dot{\phi}+\frac{dV}{d\phi}+3 H^2\left(2F_1(\phi)\ddot{\phi}+\frac{dF_1}{d\phi}\dot{\phi}^2\right)
+18 H^3F_1(\phi)\dot{\phi}+\\
&12 H\dot{H}F_1(\phi)\dot{\phi}+24\left(\dot{H}H^2+H^4\right)\frac{dF_2}{d\phi}=0
\end{aligned}
\ee
In general the couplings $F_1(\phi)$ and $F_2(\phi)$ could be arbitrary functions of the scalar field, which gives more general character to the model (\ref{eq1}) where the couplings should be constrained by known observational limits. General couplings also allow to increase the number of phenomenologically viable solutions to the DE problem.
Based on effective limits of fundamental theories like super gravity or string theory, the kinetic and GB couplings appear as exponentials of the scalar field (in leading $\alpha'$ correction in the case of string theory), but if take into consideration higher order corrections in $\alpha'$ expansion in the effective string theory, the couplings should change. Here we will consider the string inspired model with exponential couplings of the form
\be\label{eq14}
F_1(\phi)=\xi e^{\alpha\kappa\phi/\sqrt{2}},\,\,\,\, F_2(\phi)=\eta e^{\alpha\kappa\phi/\sqrt{2}},\,\,\,\, V(\phi)=V_0 e^{-\alpha\kappa\phi/\sqrt{2}}
\ee
where we also consider an exponential potential that allows to study power-law solutions.\\
In this section we will consider only the standard scalar field corresponding to $\gamma=1$. Lets propose the solution
\be\label{eq14a}
H=\frac{p}{t},\,\,\,\,\, \phi=\phi_0\ln\frac{t}{t_1}
\ee
for quintessential power-law (QPL) expansion, and
\be\label{eq14b}
H=\frac{p}{t_s-t},\,\,\,\,\, \phi=\phi_0\ln\frac{t_s-t}{t_1}
\ee
for phantom power-law expansion (PPL). By replacing (\ref{eq14a}) and (\ref{eq14b}) in the Friedmann equation (\ref{eq9}), one obtains the restrictions
\be\label{eq14c}
\frac{2\sqrt{2}}{\alpha\kappa}=\phi_0
\ee
\be\label{eq14d}
\frac{3p^2}{\kappa^2}=\frac{1}{2}\gamma\phi_0^2+V_0 t_1^2+\frac{9\xi p^2}{t_1^2}\phi_0^2\pm \frac{48\eta p^3}{t_1^2}
\ee
and from the equation of motion (\ref{eq13}) we find
\be\label{eq14e}
\gamma(\pm 3p-1)\phi_0^2-2V_0t_1^2+\frac{6\xi p^2(\pm 3p-2)}{t_1^2}\phi_0^2\pm\frac{48\eta p^3(\pm p-1)}{t_1^2}=0
\ee
where the lower minus sign follows for the phantom solution. Let's consider the two cases separately.\\
\noindent {\bf Quintessential power-law expansion}\\
From Eq. (\ref{eq14d}) and (\ref{eq14e}) for the quintessence expansion one finds $V_0$ as
\be\label{eq14f}
V_0=\frac{\big(6\xi p^2(3p-1)+\gamma(5p-1)t_1^2\big)\kappa^2\phi_0^2+6p^2(p-1)t_1^2}{2(p+1)\kappa^2t_1^4}
\ee
For $\gamma=1$, the conditions $\xi>0$ and $p>1$ are enough to give positive potential, i.e. $V_0>0$. Thus in the regime of accelerated expansion, $V_0>0$. One might consider negative kinetic coupling $\xi<0$, which leads to the following condition by demanding positivity of the potential
\be\label{eq14g}
\kappa^2\phi_0^2<\frac{6p^2(p-1)t_1^2}{(1-5p)t_1^2-6\xi p^2(3p-1)}
\ee
Assuming $\xi=-bt_1^2$ ($b>0$), this condition reduces to 
\be\label{eq14h}
\kappa^2\phi_0^2<\frac{6p^2(p-1)}{1-5p+6bp^2(3p-1)}
\ee
Taking for instance $b=1$, it is clear that in the regime of accelerated expansion ($p>1$) this condition can be satisfied.\\
\noindent On the other hand, considering the phantom model ($\gamma=-1$), then in the regime of accelerated expansion ($p>1$) the condition
\be\label{eq14hh}
\kappa^2\phi_0^2<\frac{6p^2(p-1)t_1^2}{(5p-1)t_1^2-6\xi p^2(3p-1)},\,\,\,\, 0<\xi<\frac{(5p-1)t_1^2}{6p^2(3p-1)},\,\,\text{or},\,\,\,\, \xi<0
\ee
allows quintessence PL.\\
Let's consider only the effect of the kinetic coupling (the effect of the GB coupling has been considered in \cite{sergeiod}). Setting $\eta=0$ in Eqs. (\ref{eq14d}) and (\ref{eq14e}) and solving with respect to $V_0$ and $\xi$ one finds
\be\label{eq14h1}
V_0=\frac{\gamma(6p-1)\kappa^2\phi_0^2+6p^2(3p-2)}{2(3p+1)\kappa^2t_1^2},\,\,\,\, \xi=\frac{(2p-\gamma\kappa^2\phi_0^2)t_1^2}{2p(3p+1)\kappa^2\phi_0^2}
\ee
A reasonable assumption for $\phi_0$ could be $\kappa^2\phi_0^2=1$ which leads to 
\be\label{eq14h3}
V_0=\frac{18p^3-12p^2+\gamma(6p-1)}{2(3p+1)t_1^2}M_p^2,\,\,\,\, \xi=\frac{2p-\gamma}{2p(3p+1)}t_1^2
\ee
In the region of interest that is $p>1$ (for both QPL and PPL), we can see that independently of $\gamma$ the potential is positive. Therefore we have QPL solution for the case of pure kinetic coupling ($\eta=0$). \\
A closely related to string theory is the case where $V(\phi)=0$. To cancel the potential, it follows from (\ref{eq14f}) that 
\be\label{eq14h5}
\phi_0^2=-\frac{6p^2(p-1)t_1^2}{6\xi p^2(3p-1)+\gamma(5p-1)t_1^2}M_p^2
\ee
Considering $\gamma=1$, we note that if $\xi>0$ then the sufficient condition to have $\phi_0^2>0$ is that $1/3<p<1$. In this case in absence of potential we can not have accelerated expansion. More interesting is the case when $\xi<0$. Let's represent $\xi=-b t_1^2$, where $b>0$. Then the condition to cancel the potential becomes
\be\label{eq14h6}
\phi_0^2=\frac{6p^2(p-1)}{6b p^2(3p-1)-5p+1}M_p^2
\ee
If we take for instance $b=1$, then there are two regions of the $p$-parameter line that satisfy the requirement $\phi_0^2>0$: $0.182<p<0.633$ which is applicable to early-time cosmology, and $p>1$ which gives accelerated expansion. Then in absence of potential it is possible to describe late time QPL expansion in the frame of the present model.\\
Considering $\gamma=-1$, if $\xi<0$, then from (\ref{eq14h5}) follows that for $p>1$, $\phi_0^2>0$, and if $\xi>0$, then the consistency of (\ref{eq14h5}) demands that $p>1$ and $\xi<\frac{(5p-1)t_1^2}{6p^2(3p-1)}$. 
In the particular case of the model (\ref{eq1}) without GB coupling, the condition to cancel the potential leads to
\be\label{eq14h7}
\phi_0^2=\frac{6p^2(2-3p)}{\gamma(6p-1)}M_p^2,\,\,\,\, \xi=\frac{\gamma(1-3p)t_1^2}{6p^2 (3p-2)}
\ee
in the case $\gamma=1$, the first equality has sense only for $1/6<p<2/3$ (see eq. (3.8) in \cite{granda3}), which is appropriate for early-time cosmology. In the case $\gamma=-1$, the first equality is consistent for $p>2/3$, which allows accelerated expansion for $p>1$.\\

\noindent {\bf Phantom power-law expansion}\\
For the case of PPL expansion one finds from (\ref{eq14d}) and (\ref{eq14e}) (lower sign) for $V_0$
\be\label{eq14i}
V_0=\frac{\big(6\xi p^2(3p+1)+\gamma(5p+1)t_1^2\big)\kappa^2\phi_0^2+6p^2(p+1)t_1^2}{2(p-1)\kappa^2t_1^4}
\ee
For $\gamma=1$, the potential is always positive for $\xi>0$ and $p>1$. For negative kinetic coupling, assuming for instance $\xi=-bt_1^2$ ($b>0$) one finds the condition for positive potential
\be\label{eq14j}
\kappa^2\phi_0^2<\frac{6p^2(p+1)}{6bp^2(3p+1)-5p-1}
\ee
taking $b=1$ for instance, this restriction is consistent for any $p>1$.\\
Assuming $\gamma=-1$, then for $\xi>0$ the potential is positive provided that $p>1$ and (setting $\xi=bt_1^2$, $b>0$)
\be\label{eq14j1}
\kappa^2\phi_0^2<\frac{6p^2(p+1)}{5p-6bp^2(3p+1)+1}. 
\ee
For $\xi<0$ and $p>1$ (setting $\xi=-bt_1^2$) one finds
\be\label{eq14j2}
\kappa^2\phi_0^2<\frac{6p^2(p+1)}{6bp^2(3p+1)+5p+1}
\ee
Considering only the effect of the kinetic coupling (i.e. $\eta=0$) one finds
\be\label{eq14k}
V_0=\frac{6p^2(3p+2)+\gamma(6p+1)\kappa^2\phi_0^2}{2(3p-1)\kappa^2t_1^2},\,\,\,\, \xi=-\frac{(2p+\gamma\kappa^2\phi_0^2)t_1^2}{2p(3p-1)\kappa^2\phi_0^2}
\ee
For $\gamma=1$, this potential is positive for $p>1/3$ and in this case the kinetic coupling becomes negative ($\xi<0$). Hence, in the case $\eta=0$ the PPL expansion takes place for negative kinetic coupling. For $\gamma=-1$, the potential is positive for $p>1/3$ and $\kappa^2\phi_0^2>\frac{6p^2(3p+2)}{6p+1}$.\\
It is interesting also to study the conditions to cancel the potential in the case of PPL. From (\ref{eq14i}) follows
\be\label{eq14k1}
\phi_0^2=-\frac{6p^2(p+1)t_1^2}{\gamma(5p+1)t_1^2+6\xi p^2(3p+1)}M_p^2
\ee
For $\gamma=1$, it follows that for $\xi>0$ there is not way to cancel the potential for $p>0$. But if we consider the negative coupling $\xi<0$, then (setting $\xi=-bt_1^2$, $b>0$) the condition to cancel the potential leads to
\be\label{eq14m}
\phi_0^2=\frac{6p^2(p+1)}{6bp^2(3p+1)-5p-1}M_p^2
\ee
taking for instance $b=1$, it follows that (\ref{eq14m}) is consistent for $p>0.483$. Therefore it is possible to have PPL in absence of potential. Taking $\gamma=-1$ in (\ref{eq14k1}), then for $\xi<0$ there is always possible to have PPL for any $p$, and for $\xi>0$, the restriction (\ref{eq14k1}) is consistent provided that $\xi<\frac{(5p+1)t_1^2}{6p^2(3p+1)}$.\\
If we limit the model and consider only the effect of the kinetic coupling ($\eta=0$), then from (\ref{eq14k}) for $\gamma=1$ follows that there is not way to have $V_0=0$. Hence, in order to have PPL expansion in this case, it is necessary to have a potential. In the case of $\gamma=-1$, the condition to cancel the potential leads to 
\be\label{eq14m1}
\phi_0^2=\frac{6p^2(3p+2)}{6p+1}M_p^2,\,\,\,\,\, \xi=\frac{(3p+1)t_1^2}{6p^2(3p+2)}
\ee
which is consistent for any $p>0$. Another important solution of the model (\ref{eq1}) with the potential and couplings as given by (\ref{eq14}) is the de Sitter solution. In fact, if we consider the scalar field and the Hubble parameter as constants, i.e. $\phi=const.=c$ and $H=const.=H_1$, then by replacing in  (\ref{eq9}) and (\ref{eq13}) we find
\be\label{eq14n}
H_1^2=-\frac{1}{8\eta\kappa^2}e^{-2c/\phi_0} 
\ee
Note that in this solution the kinetic coupling is irrelevant since $\dot{\phi}=0$ (the solution is possible for negative $\eta$).\\ 
In the important the case of $V(\phi)=0$, as follows from (\ref{eq14h6}) and (\ref{eq14m}) there is an asymptotic de Sitter solution at $p\rightarrow \infty$ where
\be\label{eq14o}
\phi^2=\frac{1}{3b}M_p^2
\ee
\section{Possible mechanisms to avoid the BR singularity}
The PPL solution suffers the well known problem of the future BR singularity at $t=t_s$ \cite{caldwell,nesseris,sodintsov2,bamba}. We may use the fact that in the frame of the present model, the PPL can be obtained in the absence of potential, and that in the case of dominance of potential over the other interaction terms the model presents asymptotic quintom behavior. This fact could provide a mechanism to evade the future BR singularity as follows. Let's focus on the PPL in the case when we can neglect the effect of the potential. To this end we propose the following model
\be\label{eq15}
F_1(\phi)=\xi e^{2\phi/\phi_0},\,\,\,\, F_2(\phi)=\eta e^{2\phi/\phi_0},\,\,\,\, V(\phi)=V_0 e^{-2(1+\delta)\phi/\phi_0},\,\,\,\, (\delta>0)
\ee
It is clear that the power-law is not a solution to this model, unless $\delta=0$. Nevertheless, we can make some qualitative analysis based on asymptotic behavior of the model. 
In this case, when the curvature is small we assume that the solution behaves as (\ref{eq14b}), which in absence of potential leads to the condition (\ref{eq14m}), that follows for PPL expansion. Small curvature means that $(t_s-t)$ is large, and the potential that behaves like $V\propto 1/(t_s-t)^{2(1+\delta)}$, can be neglected compared to the kinetic and GB couplings that behave as $1/(t_s-t)^2$. But as the universe evolves, the difference $(t_s-t)$ becomes smaller and the curvature increases as $t\rightarrow t_s$. At this stage the potential becomes dominant over the GB and kinetic couplings and the model becomes dominated by pure quintessential scalar field. In this case the only possible power-law solution for the potential (\ref{eq15}) is of the form
\be\label{eq16}
H=\frac{p}{t},\,\,\,\,\,\, \phi=\frac{\phi_0}{1+\delta}\ln\frac{t}{t_1}
\ee
which leads to the known conditions
\be\label{eq16a}
\phi_0^2=2p(1+\delta)^2 Mp^2,\,\,\,\, V_0=\frac{p(3p-1)}{t_1^2}M_p^2
\ee
giving an EoS parameter $w>-1$, avoiding in this way the future BR singularity. As the universe continue evolving after the dominance of the potential, the curvature turns again to small values and the interacting terms start dominating again. Nevertheless, when the potential and the couplings are present there is an important solution, namely the de Sitter solution. In fact, if we assume for the model (\ref{eq15})
\be\label{eq16a1}
\phi=const. = c,\,\,\,\, H=const.=H_1
\ee
then, by replacing in (\ref{eq9}) and (\ref{eq13}) this solution gives
\be\label{eq16a2}
H_1^2=-\frac{1}{8\eta\kappa^2}e^{-2c/\phi_0},\,\,\,\, c=\frac{\phi_0}{2\delta}\ln\left(-\frac{8\eta\kappa^4 V_0}{3}\right)
\ee
This is an alternative to the approach presented in \cite{sergeiod} where the GB coupling and phantom scalar field were considered. If we consider the phantom version of the model (i.e. with $\gamma=-1$), then there are more alternatives to avoid the BR singularity.\\
\subsection*{Phantom scalar field $\gamma=-1$}
Bellow we consider the phantom kinetic term in the model (\ref{eq1}) (i.e. assuming $\gamma=-1$), which gives more possible ways of evading the BR singularity, namely\\

{\it 1. In absence of kinetic coupling ($\xi=0$).}\\ In this case, the model reproduces the same results presented in \cite{sergeiod}.\\

{\it 2. In absence of GB coupling ($\eta=0$).}\\ One may consider the situation when initially at low curvature (large time) the potential term dominates during PPL expansion, and then when the solution is closer to the BR singularity the kinetic coupling becomes dominant, allowing the possibility of QPL expansion with EoS parameter $w>-1$. To this end we propose the model
\be\label{eq17}
V=V_0 e^{-2\phi/\phi_0},\,\,\,\,\,\, F_1=\xi e^{2(1+\delta)\phi/\phi_0},\,\,\,\, (\delta<0)
\ee
Assuming that the solution behaves as (\ref{eq14b}), then neglecting the kinetic coupling, from (\ref{eq9}) and (\ref{eq13}) (for $\gamma=-1$) follow the known conditions for existence of PPL solution (\ref{eq14b})
\be\label{eq18}
\phi_0^2=2p M_p^2,\,\,\,\, V_0=\frac{p(3p+1)}{t_1^2}M_p^2
\ee
as the curvature increases when $t\rightarrow t_s$, the kinetic coupling becomes dominating and the potential could be neglected. In this case we may assume the same solution (\ref{eq16}) for the quintessential expansion, that being replaced in (\ref{eq9}) and (\ref{eq13}) (setting $V=0$ and $F_2=0$)  leads to the conditions
\be\label{eq19}
\phi_0^2=\frac{6p^2(3p-2)}{6p-1}(1+\delta)^2M_p^2,\,\,\,\, \xi=\frac{3p-1}{6p^2(3p-2)}t_1^2
\ee
which is consistent for $p>2/3$. So there is possible to change the effective EoS from $w<-1$ to $w>-1$ avoiding the BR singularity. Hence, when the term with kinetic coupling becomes dominant, the BR singularity might be prevented. So the kinetic coupling may play a role similar to the GB coupling \cite{sergeiod} in working against the singularity.\\ 
For the power-law solution of the form (\ref{eq16}), after the dominance of the kinetic coupling the curvature becomes small again, and the potential term recovers his dominance, but there exists a solution when both, the potential and the kinetic coupling are present  which corresponds to a de Sitter phase. If we assume the solution (\ref{eq16a1}), then replacing in (\ref{eq9}) and (\ref{eq13}) with (\ref{eq17}), one finds
\be\label{eq20}
H_1^2=\frac{\kappa^2 V_0}{3} e^{-2c/\phi_0}
\ee
which is valid for arbitrary $c$, since $\dot{\phi}=0$, making this solution independent of the kinetic coupling. Therefore, there is a possibility that the universe enters in a de Sitter phase after domination of the kinetic coupling.\\

{\it 3. In absence of potential ($V=0$).}\\ 
Here we consider the following model
\be\label{eq21}
F_1=\xi e^{2\phi/\phi_0},\,\,\,\,\, F_2=\eta e^{2(1+\delta)\phi/\phi_0},\,\,\,\, \delta<0
\ee
Assuming the phantom solution (\ref{eq14b}), we see that at low curvature when $t_s-t$ is large, the GB coupling behaves as $1/(t_s-t)^{2+2\delta}$ and can be neglected with respect to the term with kinetic coupling. In this case, by solving Eqs. (\ref{eq9}) and (\ref{eq13}) one finds the conditions
\be\label{eq22}
\phi_0^2=\frac{6p^2(3p+2)}{6p+1}M_p^2,\,\,\,\,\, \xi=\frac{3p+1}{6p^2(3p+2)}t_1^2
\ee
which corresponds to phantom phase with effective EoS $w<-1$. When the curvature turns to large values at $t\rightarrow t_s$, the term with kinetic coupling could be neglected giving rise to the dominance of the GB term. Neglecting the kinetic coupling in (\ref{eq21}) and assuming a solution of the form given by eq. (\ref{eq16}), one finds from (\ref{eq9}) and (\ref{eq13})
\be\label{eq23}
\phi_0^2=\frac{6p^2(p-1)}{5p-1}(1+\delta)^2 M_p^2,\,\,\,\,\, \eta=\frac{3p-1}{8p(5p-1)}M_p^2 t_1^2
\ee
which is consistent for $p>1$, leading to quintessential expansion with $w>-1$.\\
Note that the role of the kinetic and GB couplings could be changed, i.e. one starts with domination of the GB term and ends with domination of the kinetic coupling, by proposing
\be\label{eq24}
F_1=\xi e^{2(1+\delta)\phi/\phi_0},\,\,\,\,\, F_2=\eta e^{2\phi/\phi_0},\,\,\,\, \delta<0
\ee
In this case, at low curvature the kinetic coupling can be neglected, and replacing the PPL solution (\ref{eq14b}) in (\ref{eq9}) and (\ref{eq13}) one finds the conditions
\be\label{eq25}
\phi_0^2=\frac{6p^2(p+1)}{5p+1}M_p^2,\,\,\,\, \eta=-\frac{3p+1}{8p(5p+1)}M_p^2 t_1^2
\ee
which is consistent for any $p>0$ and negative $\eta$. At large curvature when $t\rightarrow t_s$, the kinetic term becomes dominant and we can propose the solution (\ref{eq16}), which being replaced in (\ref{eq9}) and (\ref{eq13}) gives the restrictions
\be\label{eq26}
\phi_0^2=\frac{6p^2(3p-2)}{6p-1}(1+\sigma)^2M_p^2,\,\,\,\,\, \xi=\frac{3p-1}{6p^2(3p-2)}
\ee
which is consistent for accelerated expansion with $p>1$ and effective EoS  $w>-1$. However in the model without potential, there is not de Sitter solution corresponding to constant scalar field.\\

{\it 4. With all the terms.}\\ 
We may consider the phantom scalar model with the potential and couplings given by  
\be\label{eq27}
F_1(\phi)=\xi e^{2(1+\delta)\phi/\phi_0},\,\,\,\, F_2(\phi)=\eta e^{2(1+\delta)\phi/\phi_0},\,\,\,\, V(\phi)=V_0 e^{-2\phi/\phi_0},\,\,\,\, (\delta<0)
\ee
If we neglect the couplings $F_1$ and $F_2$ at low curvature, then the only possible power-law solution is the phantom one, given by (\ref{eq14b}). When the curvature becomes large at $t\rightarrow t_s$, the interacting terms become relevant and (neglecting the potential) there is a power-law solution of the form (\ref{eq16}) which leads to
\be\label{eq28}
\phi_0^2=\frac{6p^2(p-1)t_1^2}{(5p-1)t_1^2-6p^2\xi(3p-1)}(1+\delta)^2M_p^2
\ee
which is positive whenever $t_1^2>6\xi p^2(3p-1)/(5p-1)$. For negative $\xi$ this condition is satisfied for any $p>1$, leading to $w>-1$ and avoiding the BR singularity. Note that the solution (\ref{eq28}) exists in the asymptotic de Sitter limit at $p\rightarrow\infty$ ($w\rightarrow -1$), given by
\be\label{eq29a}
\phi_0^2=-\frac{(1+\delta)^2t_1^2}{3\xi}M_p^2
\ee
valid for $\xi<0$ (in this limit $\eta\rightarrow 0$).
After the dominance of the coupling terms the curvature begins to decrease again, but there exists a de Sitter solution (\ref{eq16a1}) when the three terms in (\ref{eq27}) are present
\be\label{eq29}
H_1^2=-\frac{1}{8\eta\kappa^2}e^{-2(1+\delta)c/\phi_0},\,\,\,\,\, c=\frac{\phi_0}{2\delta}\ln\left(-\frac{3}{8\eta\kappa^4 V_0}\right)
\ee
So it is possible to implement the asymptotic mechanism to avoid the BR singularity as proposed in \cite{sergeiod}, in different variants of the model depending on the correlation between the kinetic coupling, the GB coupling and the potential. As has been shown, this mechanism can be implemented in the standard and phantom version of the scalar field.\\
It is worth mentioning that the account of quantum effects near the singularity were also considered to moderate the BR singularity \cite{sdodintsov1}.
Another interesting alternative to avoid the future BR singularity is provided by the solutions known as ``Little Rip'' (LR) \cite{frampton1,sdodintsov,frampton2,granda10}, which are free of future singularity. The LR solutions produce late-time cosmological effects similar to that of the BR solutions, as the rapid expansion in the near future with an EoS $w<-1$, but the scale factor and density remain finite in finite time. As in the case of BR, the LR solutions also lead to the dissolution of all bound structures in the universe in the future.
\section{Stability of the power-law solution for the string motivated model}
We use the dynamical system approach in order to analyze the stability of the above power-law solutions, in the specific case of $V=0$, that apart from simplifying the dynamical system is also closely related to string theory. Let's introduce the following dimensionless variables:
\be\label{eq45}
\begin{aligned}
&x=\frac{\kappa\dot{\phi}}{\sqrt{2}H}, \,\,\, k=3\kappa^2F_1\dot{\phi}^2, \\
& g=8\kappa^2 H\dot{\phi}\frac{dF_2}{d\phi},\,\,\, \epsilon=\frac{\dot{H}}{H^2}
\end{aligned}
\ee
In fact, these variables are related with the density parameters for the different sectors of the model
\be\label{eq46}
\Omega_{\phi}=\frac{\kappa^2\rho_{\phi}}{3H^2}=\frac{1}{3}(x^2+y),\,\,\,\, \Omega_k=\frac{\kappa^2\rho_k}{3H^2}=k,\,\,\, \Omega_{GB}=\frac{\kappa^2\rho_{GB}}{3H^2}=g
\ee
where
\be\label{eq47}
\rho_{\phi}=\frac{1}{2}(\dot{\phi}^2+V),\,\,\, \rho_k=9\kappa^2H^2F_1\dot{\phi}^2,\,\,\,\, \rho_{GB}=24\kappa^2H^3\frac{dF_2}{dt}
\ee
The Eq. (\ref{eq9}) imply the following restriction on the density parameters
\be\label{eq48}
\Omega_{\phi}+\Omega_k+\Omega_{GB}=1
\ee
and the effective equation of state (EoS) can be written as
\be\label{eq49}
w_{eff}=w_{\phi}\Omega_{\phi}+w_k\Omega_k+w_{GB}\Omega_{GB}=-1-\frac{2}{3}\epsilon
\ee
Introducing the e-folding variable $N=\log a$, in terms of the variables (\ref{eq15}), the Eqs. (\ref{eq9})-(\ref{eq14}) can be transformed into the following first-order autonomous system
\be\label{eq50}
\gamma x^2+3k+3g-3=0
\ee
\be\label{eq51}
2\gamma x x'+2\gamma(3+\epsilon)x^2+k'+2(3+2\epsilon)k+3(1+\epsilon)g=0
\ee
\be\label{eq52}
2\epsilon +3+\gamma x^2-\frac{2}{3}k'-\frac{1}{3}(3+2\epsilon)k-g'-(2+\epsilon)g=0
\ee
\be\label{eq55}
k'=(\alpha x+2\epsilon)k+2\frac{x'}{x}k
\ee
\be\label{eq56}
g'=(\alpha x+2\epsilon)g+\frac{x'}{x}g
\ee
where ``$'$'' denotes derivative with respect to $N$ and $\gamma=\pm 1$ is the sign of the free kinetic term. Note that the last three Eqs. come from the explicit form of the potential and the couplings given in (\ref{eq14}). From Eq. (\ref{eq52}) follows the expression for the slow-roll parameter $\epsilon$
\be\label{eq57}
\epsilon=\frac{9+3\gamma x^2-2k'-3k-3g'-6g}{2k+3g-6}
\ee
It is easy to check that the power-law solutions (\ref{eq14a}) and (\ref{eq14b}) are critical points of the system, i.e., if we write the dynamical variables (\ref{eq45}) for $H$ and $\phi$ given by (\ref{eq14a}) or (\ref{eq14b}) as
\be\label{eq58}
x_0=\pm\frac{\kappa\phi_0}{\sqrt{2}p},\,\, k_0=\frac{3\xi\kappa^2\phi_0^2}{t_1^2},\,\, g_0=\pm\frac{16\eta\kappa^2 p}{t_1^2}
\ee
where the ``-'' sign is for the PPL, then these variables satisfy the equations: $x_0'=k_0'=g_0'=0$. So we will consider small perturbations
\be\label{eq59}
x=x_0+\delta x,\,\, k=k_0+\delta k,\,\,\, g=g_0+\delta g
\ee
and check the stability around the critical point $(x_0, k_0, g_0)$. Solving the system (\ref{eq50})-(\ref{eq57}) with respect to $x', k', g'$ one can write
\be\label{eq60}
x'=f_1(x,k,g),\,\,\,\ k'=f_2(x,k,g),\,\,\, g'=f_3(x,k,g)
\ee
for small perturbations, $x', k', g'$ suffer the change
\be\label{eq61}
\left( 
\begin{array}{c}
\delta x' \\ \delta k'\\ \delta g'
\end{array}
\right)=\left(
\begin{array}{ccc}
\frac{\partial f_1}{\partial x}& \frac{\partial f_1}{\partial k}& \frac{\partial f_1}{\partial g}\\ 
\frac{\partial f_2}{\partial x}& \frac{\partial f_2}{\partial k}& \frac{\partial f_2}{\partial g}\\ 
\frac{\partial f_3}{\partial x}& \frac{\partial f_3}{\partial k}& \frac{\partial f_3}{\partial g}
\end{array}
\right)\left(
\begin{array}{c}
\delta x \\ \delta k\\ \delta g
\end{array}
\right)
\ee
where the matrix is valuated at the fixed point ($x_0, k_0, g_0$) given by (\ref{eq58}). The stability under small perturbations demand that the eigenvalues of the above matrix be negative or complex with negative real component. We will analyze the stability for two different cases. In the first case we consider the model with only kinetic coupling ($g=0$), and in the second case we consider both couplings (the case with only GB coupling was considered in \cite{sergeiod}).\\

{\bf The model with non-minimally coupled kinetic term.}\\
By setting $g=0$ in (\ref{eq50})-(\ref{eq57}), it reduces to a two dimensional system for $x$ and $k$, and for small perturbations we may write 
\be\label{eq62}
\left( 
\begin{array}{c}
\delta x' \\ \delta k'
\end{array}
\right)=M \left(
\begin{array}{c}
\delta x \\ \delta k
\end{array}
\right)
\ee
where the two dimensional matrix $M$ is evaluated at the critical point $(x_0, k_0)$, which gives the components
\be\nonumber
M_{11}=\frac{3(5x_0^2\mp 9)x_0^2 -(6\alpha x_0-9)k_0^2 + 
 (\pm 72 x_0^2-6\alpha x_0\mp 8\alpha x_0^3+9)k_0}{
6k_0^2\pm 6x_0^2-2(\pm x_0^2-3)k_0}
\ee
\be\nonumber
M_{12}=\frac{(\pm 24x_0^2-3\alpha x_0\mp 2\alpha x_0^3-6(\alpha x_0-3)k_0+9)x_0}{6k_0^2\pm 6x_0^2-2(\pm x_0^2-3)k_0}
\ee
\be\nonumber
M_{21}=\mp \frac{3((\alpha x_0-8)k_0-3\alpha x_0+12)k_0 x_0}{
 3k_0^2\pm 3x_0^2-(\pm x_0^2-3)k_0}
\ee
\be\label{eq63}
M_{22}=\pm \frac{(24k_0+3\alpha x_0-2\alpha k_0 x_0-18)x_0^2}{
3k_0^2\pm 3x_0^2-(\pm x_0^2-3)k_0}
\ee
where the lower sign is assigned to the phantom model ($\gamma=-1$, see Eqs. (\ref{eq9}) and (\ref{eq13})). Replacing $x_0$ and $k_0$ from (\ref{eq58}) and using the restrictions (\ref{eq14h7}) for QPL, we find the following eigenvalues for the matrix $M$
\be\label{eq64}
\lambda_1=\frac{1-3p}{p},\,\,\ \lambda_2=\frac{2-3p}{p},\,\,\,\, \frac{1}{6}<p<\frac{2}{3}
\ee
where $p$ should obey the above restriction for consistency, according to (\ref{eq14h7}). In the interval $1/6<p<2/3$, $\lambda_2>0$ and therefore the power-law solution of the form $H=p/t$ is unstable for the model in absence of potential and GB term. Note that the PPL solution with only kinetic coupling is not possible as was demonstrated above (see eq. (\ref{eq14k}) for $V_0=0$).\\
On the other hand, if we consider the phantom model that obeys the Eqs. (\ref{eq9}) and (\ref{eq13}) ($\gamma=-1$), then taking into account the lower sign in the components (\ref{eq63})
one finds the following eigenvalues for the QPL
\be\label{eq65}
\lambda_1=\frac{1-3p}{p},\,\,\,\, \lambda_2=\frac{2-3p}{p},\,\,\,\, p>\frac{2}{3}
\ee
where the last inequality follows from the consistency of the solution of Eqs. (\ref{eq9}) and (\ref{eq13}) with $V=0$ and $\eta=0$, for QPL (see eq. (\ref{eq19})). Therefore, for the phantom model with kinetic coupling, the power-law solution (\ref{eq14a}) is a stable fixed point provided $p>2/3$.\\
If we consider the phantom model with PPL (\ref{eq14b}), then using the lower sign in (\ref{eq63}) and in (\ref{eq58}) one finds the following eigenvalues
\be\label{eq66}
\lambda_1=-\frac{3p+2}{p},\,\,\,\, \lambda_2=-\frac{3p+1}{p}
\ee
which is valid for any $p>0$, as follows from Eqs. (\ref{eq9}) and (\ref{eq13}) with $V=0$ and $\eta=0$, for PPL. Then the power-law solution (\ref{eq14b}) is a stable fixed point for the phantom model with $V=0$ and $\eta=0$.\\

\noindent {\bf The model with kinetic and GB couplings}\\
Here we consider the stability of power-law solution for the three dimensional autonomous system (\ref{eq50})-(\ref{eq57}) in the following cases: \\
{\it Quintessence and phantom power-law for $\gamma=1$.}\\
Evaluating the matrix elements of eq. (\ref{eq61}) for the fixed point (\ref{eq58}) (upper sign) and for the QPL (upper sign), we find the following eigenvalues
\be\label{eq67}
\lambda_1=0,\,\,\,\, \lambda_2=\frac{1-3p}{p},\,\,\,\,\, \lambda_3=\frac{2-3p}{p}
\ee
note that $\lambda_2$ and $\lambda_3$ are negative for $p>2/3$. This restriction is compatible with the condition of consistency for the QPL solution (\ref{eq14a}), as can be seen in (\ref{eq14h6}). \\
For the PPL we have found the eigenvalues
\be\label{eq68}
\lambda_1=0,\,\,\,\, \lambda_2=-\frac{2+3p}{p},\,\,\,\,\, \lambda_3=-\frac{1+3p}{p}
\ee
where $\lambda_2$ and $\lambda_3$ are negative for any $p>0$, which is compatible with the restrictions
{\it Quintessence and phantom power-law for $\gamma=-1$.}\\
Evaluating the matrix elements in (\ref{eq61}) for the fixed point (\ref{eq58}) (lower sign) and for the QPL (upper sign), we find the following eigenvalues
\be\label{eq69}
\lambda_1=0,\,\,\,\, \lambda_2=\frac{1-3p}{p},\,\,\,\,\, \lambda_3=\frac{2-3p}{p}
\ee
which is compatible with the solutions in absence of potential for $p>1$ (see \ref{eq28}).\\
For PPL we have the eigenvalues 
\be\label{eq70}
\lambda_1=0,\,\,\,\, \lambda_2=-\frac{2+3p}{p}, \,\,\,\,\, \lambda_3=-\frac{1+3p}{p}
\ee
valid for $\xi<0$ and $p>0$. Note that the eigenvalues are independent of $\gamma$. So in the presence of both couplings the stability of the power-law solution does not depend on the standard or phantom character of the model. But a problem appears in both cases due to the presence of zero eigenvalues. In this case the linear expansion fails to provide information on the stability of the fixed point. We need to consider higher order corrections to study the stability of perturbations along the zero eigenvalue direction.\\

\noindent {\bf The centre manifold analysis.}\\
To analyze the stability in the presence of zero eigenvalues we use the approach of the central manifold \cite{rendall}, \cite{koivisto}, \cite{ruth}, which reduces the dimensionality of the system near the critical point, and limits the stability analysis to the reduced system. Thus the stability properties of the system become determined by the (in)stability of the reduced system. To this end we need to translate the fixed point (\ref{eq58}) to the origin, by introducing the variables (we keep the same symbols)
\be\label{eq71}
x\rightarrow x-x_0,\,\,\,\, k\rightarrow k-k_0,\,\,\,\,\, g\rightarrow g-g_0
\ee
The simplest case takes place for one zero eigenvalue, which leads to one-dimensional reduced system. Composing the matrix  $M_0$ with the eigenvectors of the Hessian in the new defined fixed point $(0,0,0)$, we introduce a new set of coordinates
\be\label{eq72}
\left( 
\begin{array}{c}
u \\ v \\ w
\end{array}
\right)=M_0 \left(
\begin{array}{c}
x \\ \ k \\ g
\end{array}
\right)
\ee
In these coordinates the dynamical equations can be written in the form
\be\label{eq73}
\left( 
\begin{array}{c}
u' \\ v'\\ w'
\end{array}
\right)=\left(
\begin{array}{ccc}
0& 0& 0\\ 
0& \frac{1-3p}{p}& 0\\ 
0& 0& \frac{2-3p}{p}
\end{array}
\right)\left(
\begin{array}{c}
u \\ v\\ w
\end{array}
\right)+
\left(\begin{array}{c}
\tilde{f_1}(u,v,w) \\ \tilde{f_2}(u,v,w)\\ \tilde{f_3}(u,v,w)
\end{array}\right)
\ee
where the last column represents the non-linear terms. Note that in our case the variable $u$ is actually the same variable $x$. Then the system can be written as
\be\nonumber
u'=\tilde{f_1}(u,y),
\ee
\be\label{eq74}
y'=L y+\tilde{f}(u,y)
\ee
where 
\be\label{eq75}
y=\left(
\begin{array}{c}
 v\\ w
\end{array}\right),\,\,\, 
L=\left(
\begin{array}{cc}
\frac{1-3p}{p}& 0\\ 
0& \frac{2-3p}{p}
\end{array}
\right),\,\,\,\, 
\tilde{f}(u,y)=\left(\begin{array}{c}
\tilde{f_2}(u,y)\\ \tilde{f_3}(u,y)
\end{array}\right)
\ee
We now turn to the definition of centre manifold:\\
The space
\be\label{eq76}
W^{c}(0)=\left\{(u,y)\in R^{1}\times R^{2}| y=h(u),|u|<\delta, h(0)=0, Dh(0)=0\right\}
\ee
where $Dh$ is the matrix of first derivatives of the vector valued function $y=h(x)$, is called the centre manifold for the system (\ref{eq74}).
Since $y=h(x)$, the dynamics of the system becomes reduced to the centre manifold in the neighborhood of $x$, and the stability properties of the full dynamical system depend on the analysis in the centre manifold. Using $y=h(x)$, the system (\ref{eq74}) leads to
\be\label{eq77}
Dh(u)\tilde{f_1}(u,h(u))=Lh(u)+\tilde{f}(u,h(u))
\ee
This differential equation can be used to find $h(u)$, and then by replacing $h(u)$ into the first equation (\ref{eq74}) (i.e. $u'=\tilde{f_1}(u,h(u))$) we can analyze the stability of the reduced system. Near the critical point we can Taylor expand $h(u)$ in powers of $u$ and calculate the coefficients of the first non-trivial terms from (\ref{eq77}). In our case we assume $h$ of the form
\be\label{eq78}
h(u)=\left(
\begin{array}{c}
 a_2 u^2+a_3 u^3+a_4 u^4+a_5 u^5+{\cal O}(u^6)\\ b_2 u^2+b_3 u^3+b_4 u^4+b_5 u^5+{\cal O}(u^6) 
\end{array}\right)
\ee
Using the restrictions on $\xi$ and $\eta$ for the QPL solution in absence of potential, the critical point (\ref{eq58}) takes the form
\be\label{eq79}
x_0=\frac{2}{\alpha p},\,\,\, k_0=-\frac{3\alpha^2 p^3-3\alpha^2 p^2+20p-4}{\alpha^2 (3p-1)p^2},\,\,\, g_0=\frac{2(9\alpha^2 p^3-6\alpha^2 p^2+24p-4)}{3\alpha^2(3p-1)p^2}
\ee
where $\alpha$ is given by (\ref{eq14c}). To avoid large analytical expressions we will limit the analysis to specific values of $p$. The first interesting value of $p$ corresponds to the de Sitter limit, which gives the critical point $(x_0=0,k_0=-1,g_0=2)$ and the corresponding eigenvalues from (\ref{eq67}) are $(\lambda_1=0,\lambda_2=-3,\lambda_3=-3)$. Applying the above central manifold analysis, and after large but straightforward calculations we find the following equation for the reduced system
\be\label{eq80}
u'=\left(\frac{9}{16\alpha^2}-\frac{3}{64}\right)u^7+{\cal O}(u^8) 
\ee
by replacing $h(\alpha)=\frac{9}{16\alpha^2}-\frac{3}{64}$ and integrating this equation one obtains
\be\label{eq80a}
u=u_0\left(1-6u_0^6h(\alpha)N\right)^{-1/6}
\ee
where $u_0$ is the initial perturbation along the zero eigenvalue direction. In order for this initial perturbation to decay it follows from (\ref{eq80a}) that independently of the sign of $u_0$, the coefficient $h(\alpha)$ must be negative. Therefore if $h(\alpha)<0$ the critical point will be stable.
Turning to eq. (\ref{eq80}) we see that if $\alpha^2>12$ then $h<0$ and the critical point is a stable attractor.\\
Another reasonable value is $p=200/9$, which gives the EoS parameter $w=-0.97$. In this case the critical point is $(x_0=\frac{9}{100\alpha}, k_0=-\frac{191}{197}-\frac{26757}{1970000\alpha^2}, g_0=\frac{388}{197}+\frac{10719}{985000\alpha^2})$ and the eigenvalues are $(\lambda_1=0,\lambda_2=-591/200,\lambda_3=-291/100)$. 
The centre manifold analysis gives 
\be\label{eq81}
u'=\mu(\alpha) u^7+{\cal O}(u^8) 
\ee
but the analytical expression for $\mu(\alpha)$ is too large to be displayed here. We limit ourselves to numerical evaluation for some values of $\alpha$: ($\alpha=2$, $\mu=0.809$), ($\alpha=3$, $\mu=0.03526$), ($\alpha=12$, $\mu=-0.03592$) and ($\alpha=14$, $\mu=-0.03549$). Note that for the last two values of $\alpha$ the critical point is stable.
\\
One can also perform the same analysis for the PPL (lower sing in (\ref{eq58}) with eigenvalues given by (\ref{eq68}). For the specific value of $p=80/3$ giving the EoS parameter $w=-1.025$, the centre manifold analysis gives  
\be\label{eq82}
u'=\nu(\alpha) u^7+{\cal O}(u^8) 
\ee
evaluating $\nu(\alpha)$ for some values of $\alpha$, gives:($\alpha=1$, $\nu=0.055$), ($\alpha=3$, $\nu=-0.008$), ($\alpha=12$, $\nu=-0.049$), ($\alpha=14$, $\nu=-0.051$). The last three values of $\alpha$ lead to stability.
\section{Discussion}
We studied late time power-law cosmological solutions based on string spired scalar-tensor model including a coupling to the Gauss-Bonnet invariant and kinetic couplings to curvature. The model allows quintessential and phantom power-law expansion in a variety of scenarios that involve different asymptotic limits. In the case with potential the conditions (\ref{eq14f}) (\ref{eq14g}) and (\ref{eq14hh}) allow QPL expansion, and the conditions (\ref{eq14i}), (\ref{eq14j}), (\ref{eq14j1}) and (\ref{eq14j2}) allow PPL expansion. In  absence of the GB coupling, the restrictions (\ref{eq14h1}) allow QPL and the restrictions (\ref{eq14k}) allow PPL. When we neglect the potential (which is closely related to string theory), the condition (\ref{eq14h5}) allows QPL expansion and (\ref{eq14k1}) allows PPL. If in addition to the potential we neglect the GB term, then the model continues to have power-law solutions, where in this case for $\gamma=1$, the condition (\ref{eq14h7}) allows power-law but in the range of decelerated expansion. And for the case of $\gamma=-1$, it follows from (\ref{eq14h7}) that the model has QPL solution. Concerning PPL solutions in this limit (i.e. $V=0$, $\eta=0$), then from (\ref{eq14k}) for $\gamma=1$ it follows that there is not PPL without potential, and in the case of $\gamma=-1$, the condition (\ref{eq14m1}) allows PPL for any $p>0$.\\ 
The model also exhibit de Sitter solution in various scenarios: considering $\phi=const.$ and $H=const.$, then from (\ref{eq9}) and (\ref{eq13}) follows the solution (\ref{eq14n}) and in absence of potential the asymptotic de Sitter solution (\ref{eq14o}) takes place. We also investigated the possible mechanism to avoid the Big Rip singularity in the case of PPL expansion. To this end we made a qualitative analysis, by proposing different asymptotic scenarios where one or two interacting terms (including the potential) in (\ref{eq1}) are relevant at low curvature, while the remaining terms become relevant at large curvature (when $t\rightarrow t_s$), providing a quintessential solution and evading in this way the singularity (see \cite{sergeiod}). After that, the universe might evolve asymptotically towards a de Sitter solution as was shown in different scenarios. The mechanism was implemented in the standard and phantom version of the scalar field (i.e. $\gamma=\pm 1$). Of special interest is the first case (\ref{eq15}) ($\gamma=1$) where at $t\rightarrow t_s$ the dynamics becomes dominated by pure quintessential scalar field, which is the only possibility.\\
We have performed the stability analysis for the string inspired model (i.e. in absence of potential) and have found that the power-law solution (\ref{eq14a}) (or (\ref{eq14b})) is a critical point of the model and is stable fixed point in different scenarios: we first considered the case where the GB coupling is neglected (i.e. $V=0$, $\eta=0$). In this case, for $\gamma=1$, the critical point is unstable in the allowable range of $p$ ($1/6<p<2/3$, see (\ref{eq14h7})). For $\gamma=-1$, the critical point (\ref{eq14a}) is stable attractor, provided that $p>2/3$. The PPL solution (\ref{eq14b} is a stable attractor for the phantom model ($\gamma=-1$) for any $p>0$. The scalar field with GB correction (i.e. with $V=0$, $\xi=0$) was considered in \cite{sergeiod}. Next we analyzed the stability of the model with GB and kinetic coupling terms. In the model with $\gamma=1$, for QPL we have found the eigenvalues $(0,(1-3p)/p,(2-3p)/p)$ and for PPL ($0,-(2+3p)/p,-(1+3p)/p$). For $\gamma=-1$, we have found the same eigenvalues. Due to the presence of zero eigenvalues, the linear expansion fails to provide information on the stability of the fixed point. We used the centre manifold analysis 
to determine the stability properties of the critical point, and found that in the important limit of de Sitter solution the critical point is a stable attractor, under certain restriction coming from the expression (\ref{eq80}). Numerical evaluation was also done for concrete values of $p$ and it was found stability for some cases.\\
Resuming, the presence of the kinetic coupling besides the GB coupling, extends the number of possible scenarios to realize cosmological solutions with BR singularity, compared to the model with only GB correction. Additionally, the different asymptotic cases of the present model not only extend such possibilities, but also provide a number of alternatives to avoid the BR singularity leading to an universe that might evolve towards a de Sitter phase. Of special interest is the first case given by the model (\ref{eq15}), in which the BR solution is obtained without appealing to phantom field (i.e. $\gamma=1$), and near the singularity the dynamics becomes dominated by purely quintessential scalar field with EoS $w>-1$.
Note that in this scenario the terms with couplings produce the PPL and the scalar potential acts against the BR singularity, while in the other cases ($\gamma=-1$, see Eqs. (\ref{eq17}), (\ref{eq21}), (\ref{eq27})) the GB and kinetic couplings might prevent the BR singularity. The above results show that the string effects could play significant role in late time cosmology.

\section*{Acknowledgments}
This work was supported by Universidad del Valle under project CI 7883.

\end{document}